\RequirePackage{fix-cm}
\documentclass[smallcondensed]{svjour3}     
\smartqed  

\usepackage{graphicx}

\usepackage{array}
\usepackage{mdwtab}
\usepackage{algpseudocode}

\usepackage{algpascal}
\usepackage{algorithm}

\usepackage[T1]{fontenc}

\usepackage[table,xcdraw]{xcolor}
\usepackage{color,soul}

\usepackage[utf8]{inputenc}

\usepackage{amsthm, amsmath, amssymb}
\usepackage{moreverb}
\usepackage{algorithm}
\usepackage[tight,footnotesize]{subfigure}

\begin{document}

\title{A Data as a Service (DaaS) Model for GPU-based Data Analytics
}


\author{John Olorunfemi Abe        \and
        Burak Berk \"{U}st\"{u}nda\v{g} 
}


\institute{ \at
             Istanbul Technical University \\
             Department of Computer Engineering, Maslak, Istanbul, Turkey\\
                           \email{joabe@itu.edu.tr, bustundag@itu.edu.tr}
}

\date{Accepted: 23 December 2017, by the IEEE/IFIP NTMS Workshop on Big Data and Emerging Trends (WBD-ET'2018); it was later withdrawn because of funding issues. An extended/enhanced version will be published in future dates in related journals }

\maketitle

\begin{abstract}
Cloud-based services with resources to be provisioned for consumers are increasingly the norm, especially with respect to Big data, spatiotemporal data mining and application services that impose a user's agreed Quality of Service (QoS) rules or Service Level Agreement (SLA). Considering the pervasive nature of data centers and cloud system, there is a need for a real-time analytics of the systems considering cost, utility and energy. This work presents an overlay model of GPU system for Data As A Service (DaaS) to give a real-time data analysis of network data, customers, investors and users' data from the datacenters or cloud system. Using a modeled layer to define a learning protocol and system, we give a custom, profitable system for DaaS on GPU. The GPU-enabled  pre-processing and initial operations of the clustering model analysis is promising as shown in the results. We examine the model on real-world data sets to model a big data set or spatiotemporal data mining services. We also produce results of our model with clustering, neural networks' Self-organizing feature maps (SOFM or SOM) to produce a distribution of the clustering for DaaS model. The experimental results thus far show a promising model that could enhance SLA and or QoS based DaaS.
\end{abstract}

\label{Introduction}

\section{Introduction}
The emergence of machine learning algorithms to enable clustering and learning of Big Data generated from the network or enterprise cloud system and other agricultural based data is of more importance at this time of research \cite{7457602,lee2015geospatial,Kurasova2014}; we have need of a real time data analytics model. Data analytics combined with a robust GPU system, in recent years, have emerged as a viable low cost, rapid service delivery model that extends the vision of Data As A Service (DaaS) model capabilities through the use of inherent framework. The real time service is with considerable challenges of updates indexes, latency issues and service oriented architecture (SOA) compliance \cite{Assuncao20153,Zhong:2013:TGL:2568486.2568513,Campa2014354,Grossman:2008:DMU:1401890.1402000,steiner2012network} .  

With Facebook, YouTube and other web based data pro- ducing enabled services for users in the tune of 50 Terabytes, a GPU-based DaaS model is for efficiency and performance with SLA and QoS with research activities in mind. Networks or systems' intelligence could be derived from the learning parameters and models of data. The learning could be done using clustering, which is a form of unsupervised or supervised learning. While research has been on CPU models for machine learning or clustering for big data in networks or enterprise system, there are very high opportunity in Data analytics for DaaS through the use of the increasing capabilities of hardware and software products. The parallelism of
GPU using CUDA, with clustering techniques could provide a model solution for this real time service. The combination of these characteristics is increasing and in readiness to compete and provide considerably solutions, but we need a middleware that will also unify all this solution.  This system should effectively use the ever increase of data in the tune of  thousands of Terabytes of service and optimizing. Modeling the workloads that exists in the Big Data Analytics scenario is one of the emerging research topics in Big Data \cite{Sondhi:2014:ALE:2616498.2616525,Azvine:2006:RTB:1153166.1153248,Xu:2015:MRT:2737182.2737186,zheng2013service,Zhong:2013:TGL:2568486.2568513}.

We provide solutions model for Data As a Service (DaaS) with respect to clustering of data in GPU and the model for optimal solutions. The process of cross correlation available through the high volume of Data on the Internet and user-generated data makes data mining a more viable part of our daily life and decision making system. Despite the availability of Big Data from Cloud networks and enterprise networks; there is a need for  real-time analytics on the GPU \cite{7416982,DeVries:2015:PSS:2736277.2741111,6650931,7056063}. The GPU instances and memory in the emerging system needs not only be memory management but also a high level middleware that can synchronize GPU operation and computation like a GPU-based DaaS. This work proposes a sort of Data As A Service (DaaS) model that possess a set of differentiation and analytic model for Big Data in different scenarios of e-health, Cyber Security, Disaster Management, spatiotemporal data mining services and Smart City.  We use a set of classified GPU components as TegraX2 and FPGA on Ubuntu servers as a test-bed \cite{steiner2012network,Vatsavai:2012:SDM:2447481.2447482,DBLP:journals/corr/abs-1208-5654,Wang:2011:FAI:1998582.1998605,6650931,7056063}.

With Moore's law being fulfilled in the power of the processors' powerful and enabled software which is basically in high growth with respect to the research of new high speed chip processors and relatively cheap with respect to speed; then the GPU systems have an advantage not just to give speed and parallelism but also high processing power of Big Data analytics. This gives an advantage for real time analysis of Big Data especially in the scenarios of e-health, cyber-security, disaster management and other related areas \cite{7457602,7140733,Assuncao20153,Azvine:2006:RTB:1153166.1153248}.

\begin{figure*}
	\centering
	\includegraphics[scale=0.7]{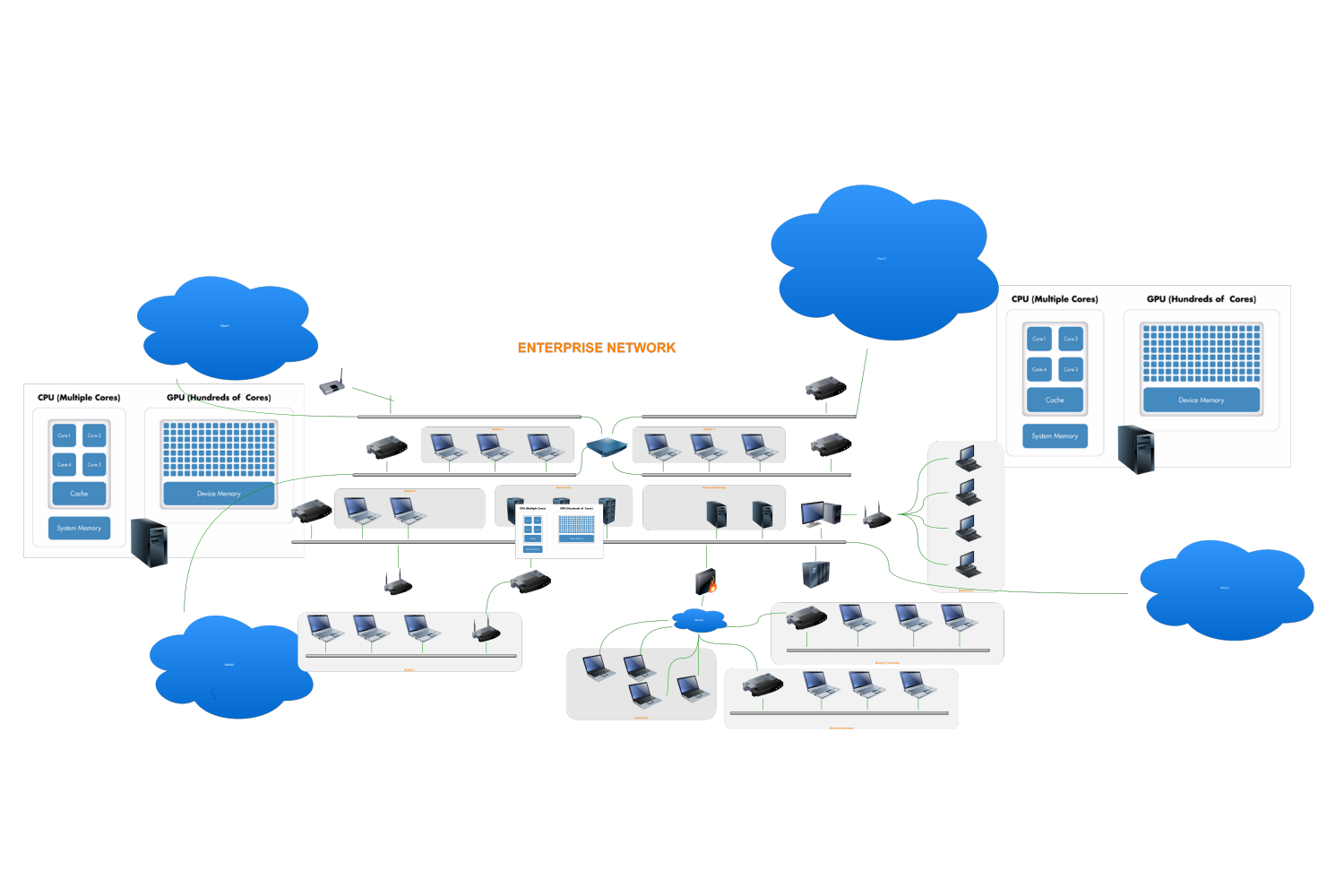}
	\caption{Data Center, Cloud and inherent GPU system- a model }
	\label{fig:Data Center, Cloud and GPU sytem- a model}
\end{figure*}

\label{contributions}
Concisely, we make the following contributions:\label{contributions paragraph}
\begin{itemize}
	
	\item Proposed a model system that will be scalable and functional for DaaS through a GPU based, machine learning enabled analytics model.
	\item Conduct preliminaries experiment on GPU enabled system to test the pre-processing system of the DaaS model.
	\item Show SOM that our model could be effectively clustered and analysis done with respect to parameters of values.	
	
\end{itemize}

\section{Related Works and Motivation }

The work in \cite{Zhong:2013:TGL:2568486.2568513} shows how clusters and cloud system of
The GPU is used to produce a fast data-intensive application, especially on the Amazon EC2 virtual cluster. The model is for an API that does threading parallelism with an in-memory processing capabilities. A graph partitioning system was proposed that also helps in load balancing of virtual cluster of the cloud systems. Our work proposes to do load-balancing by using a middleware to streamline the request for clusters of servers. The work uses CUDA abstraction that involves GPU threads that are in "warp" scheduling unit of the GPU. The graph processing capabilities is not to only do graph partitioning, but to also store a persistent storage of the cluster for future use. The GPU as a slave worker is used to execute multi-threaded Pthread and CUDA programs.

The work in \cite{5170921}  shows how the K-means algorithm are used in GPU architecture by offloading the intensive part of the operating system with programming by compute unified device architecture (CUDA). A k-means computation, analysis was done on GPU through offloading and they compared with
CPU based k-means algorithms. In \cite{Vatsavai:2012:SDM:2447481.2447482} we have Big Data in the spatiotemporal form of data and mining is done through the mixture model form with the Gaussian Mixture model (GMM) clustering algorithm. Since we have a two subproblem in this case, our modified K-means with low complexity is of a better option but a good analysis of the learning process is also needed. 

In \cite{Sondhi:2014:ALE:2616498.2616525}, e-discovery for Big Data is analyzed for advanced metadata extraction and to do prediction through respective algorithms. Since we have a high volume heterogeneous dataset of 10 Terabyte scale and the migration of present day software solutions is not easily achieved for optimal and cutting-edge platform, a high scale analytics of prediction is proposed that is implemented on High Performance Computing (HPC) systems. The challenges of storage, analytics and visualization with respect to such a Big Data also pose some good opportunities for research areas.

\section{Motivation, Proposed model and Analysis }

\subsection{GPU and CUDA }
The process of data for opportunistic scenario of various kinds has made graphics processing units (GPUs) more versatile to produce value for money. It is a processor's technology that enables parallel programming so as to analytics at real time and as at when needed system. The GPU is used to enable optimized performance. With CUDA parallel capabilities, we could do in-memory operation for GPU programming in the future \cite{Buck:2004:BGS:1186562.1015800,Campa2014354,Sondhi:2014:ALE:2616498.2616525,Wang:2011:FAI:1998582.1998605}.

Data work set at each stage of analysis of learning and process need to be modeled so that a useful amount is in-memory operation. But this is with a cost with respect to GPU chips and hardware resources either in the data centre or cloud systems. Our GPU model is to provide an offset of cost by providing a scalable virtual plane for analytic process which most importantly is also nearer to the data. We could effectively use a simple rule of association in a system that for a GPU resource being used and can deduce  costs associated with Servers, Cloud systems, and other parameters of concerns in the systems as shown below 

\begin{equation}
{\text{IF}}\;({\text{GPU Systems}}\;{\text{AND}}\;{\text{Cloud and Servers}})\; \rightarrow \;{\text{Associated Costs}}
\end{equation}

\begin{equation}{\text{IF}}\;({\text{X}}\;{\text{is}}\;{GPUm_{A}}) \rightarrow (\text{Y}\;\text{is}\;{GPUm_{B}})
\end{equation}

For the model analytics of data with SLA-enabled service, we have model and analysis and then an  interpretation system so as to produce Data Service with maximum benefit as needed \cite{zheng2013service}. For a preliminary test bed model, we use EM-tree clustering algorithm as employed in \cite{DeVries:2015:PSS:2736277.2741111,Kurasova2014} to provide a structured model for the GPU-enabled DaaS  and run on a GPU system to test our basic model. The algorithm is a basic combination of K-tree and some form of B-tree to do cluster validation and produce immutable tree state at each iteration. Using the basic K-means algorithm since it exhibits fast convergence and a high level of linear time complexity. The internal nodes of the EM-tree algorithm are kept in memory so as to produce an optimized process and iteration are of the m-order in the depths. This enables large dataset to be run on a memory intensive GPU model and produce better results. Since the K-means algorithm has a low level complexity \cite{Kurasova2014} and some level of iterations can be involved with respect to the iterations and optimization model is proposed as formulated below. The parallelized advantage of K-means is also a positive criteria on a GPU-enabled clusters or server node.

We formulated the DaaS service provisioning problem also as a continuous multi-objective optimization problem (MOP) since we have a set of objective functions and one cannot have a specific service function ${f(x)}$ to optimize the users' SLA or required QoS \cite{7457602,7416982,steiner2012network}

The model of data service demand received with SLA can be modeled as as in \cite{7416982,6319170,7056063} where a set of availability of a Data service or datacentre $	A_{VDC1}$ is modeled in first dimension as we have shown in our own adapted version as below:

\label{eq:availability of a Data service}
\begin{equation}
\begin{aligned}
A_{VDC1}=\sum\limits_{i=1}^{n}P(c_{i})A_{VDC}^{c_{i}}. 
\end{aligned}
\end{equation}
where $P(c_{i})$ is probability associated with the cost $ c_{i}$ of running the service required and probability function for the availability of the datacenters and servers is as below :

\label{eq:availability of a Data service--b}
\begin{equation}
\begin{aligned}
& \underset{x}{\text{minimize}}
& & P_\Omega(x) \\
& \text{subject to}
& & f_i(x) \leq b_i, \; i = 1, \ldots, m.
\end{aligned}
\end{equation}

$P_\Omega(x)$ is the summation of all the cost probability which subject to various set of functions related to cloud environments. Of a note is that $ P_\Omega(x) $ is possible to take upon a multiple objective function.

\label{Algorithm 1}
\hskip -0.4cm\ \begin{tabular}[t]{ p{7.5cm} }
	\hline
	\textbf{Algorithm 1:}DaaS-GPU-based model\\
	\hline
	1\hspace{0.5em}\textbf{for} {\it each \textunderscore Data \textunderscore demand f} $\in$ {\it $F_dr$} \textbf{do}\\
	2 \hspace{1.5em}input Pre-process-function({\it f});\\
	3 \hspace{1.5em}Update \textunderscore Complete \textunderscore plane \textunderscore Data({\it D,$\epsilon$});\\
	4\hspace{0.5em}\textbf{end}\\
	5\hspace{0.5em}\textbf{for} {\it each data Service f} $\in$ {\it $F_a$} \textbf{do}\\
	6\hspace{1.5em}{\it $G_K$} $\leftarrow$ Opt\textunderscore K-means\textunderscore{\it $F_r$}\textunderscore $Refine_D$ ({ \it $F_dr$,f,k});\\
	7\hspace{1.5em}{\it DaaS} $\leftarrow$ Select\textunderscore Best\textunderscore Set\textunderscore Data({\it S});\\
	8\hspace{2.7em}\textbf{if} (${\it DaaS} \neq \emptyset$) \textbf {then}\\
	9\hspace{3em}Accommodate({\it DaaS});\\
	10\hspace{3em}Update({\it\textunderscore Complete \textunderscore plane \textunderscore Data,$\epsilon$});\\
	11\hspace{2.2em}\textbf{else}\\
	12\hspace{3em}\textbf{return }  $Infeasible Solution \textunderscore Data $;\\
	13\hspace{2.2em}\textbf{end}\\
	14\hspace{0.5em}\textbf{end}\\
	15\hspace{0.6em}\textbf{return }  $Total \textunderscore optimized \textunderscore Data $;\\
	\hline\\
	
\end{tabular}

\subsection{Self-organizing Maps (SOM) }
\begin{figure}[hbtp]
	\centering
	\includegraphics[scale=0.2]{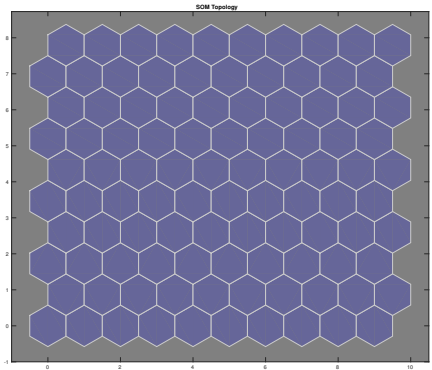}
	\caption{SOM Neighbour weight distances }
	\label{fig:SOM Neighbour weight distances}
\end{figure}

\begin{figure}[hbtp]
	\centering
	\includegraphics[scale=0.19]{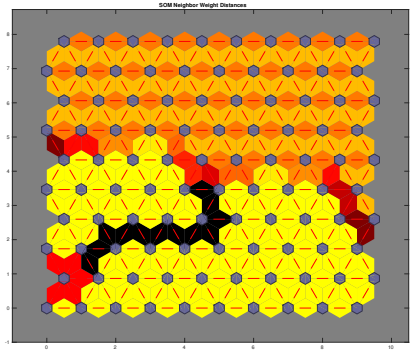}
	\caption{Model for ANN SOM Neighbor Weight Distances 200 iterations  }
	\label{fig:Model for ANN SOM Neighbor Weight Distances 200 iterations .}
\end{figure}

An analysis of the DaaS model is also done using the SOM analysis so as to provide a QoS based data service as in \cite{Mishra2012}, this comprises of a combination of modified K-means clustering and SOM. The results Figures \ref{fig:SOM Neighbour weight distances}, \ref{fig:SOM networks with hextop and gridtop topologies_but not tritop or randtop}, and  \ref{fig:Model for ANN SOM Neighbor Weight Distances 200 iterations .} as shown gives a promising distribution of data service clustering and effective SLA enabled. The The interesting differentiation of data service is shown  The equations above are to show a sort of the model approach for mathematical formulation of DaaS on GPU that we follow so as to obtain optimal efficiency

\begin{table}[thbp]
	\vskip\baselineskip 
	\caption[SOM Training Parameters]{SOM Training Parameters.}
	\begin{center}
		\begin{tabular}{|c|c|}\hline
			\textbf{Parameter}& \textbf{Value}\\\hline
			Map Size &   20 x 20 or 30 x 30\\\hline
			Lattice &   Hexagonal\\\hline
			Number of iterations & 200 \\\hline
			Training neighborhood radius  &   8 to 2\\\hline 
			Neighborhood function  & Gaussian \\\hline
		\end{tabular}
		\label{table:SOM Training Parameters' Table}
	\end{center}
\end{table}

\subsection{The GPU and CUDA programming model }

With the ubiquitous systems of GPUs, that have powerful Single Instruction Multiple Data (SIMD) processors and the accompany CUDA, a paralleled process of instructions for the model is proposed.

The push and pull model of publishing data as in \cite{Wang:2011:FAI:1998582.1998605,Eisenhauer:2006:PHC:1110639.1110693} is used so as to act as the middleware for a more robust system. This gives an advantage of a virtual real-time for data analytics. The awareness created by this model will be utilized to not the inconsistencies in data and do a real-time update. Using the K- means clustering model \cite{macqueen1967some} of data as in \cite{DBLP:journals/corr/abs-1208-5654}, we proposed a modified K- means pre-preprocessed clustering for data from the network, data and cloud centres.  With respect to the K-means clustering model problem of initial centers below equation gives a function of utilization of GPU clusters which is the part of the cost function as in \cite{Eisenhauer:2006:PHC:1110639.1110693,7140733,Zhong:2013:TGL:2568486.2568513}  that we want to optimize. The equations provided are the basic sets that formulates the model for minimizing the GPU enabled layer for analytics.

Considering a data service model that arrival is Markovian and exponential service time with a server system as an $M/M/1$, and as used in \cite{6319170} we have GPU computing data distribution equation as below: 

\label{eq:data service distribution}

\begin{eqnarray} {Objective}:   {\rm Max}\sum\limits_{i=1}^{m}\lambda_{i}b_{i}(1-e^{(\lambda_{i}-n_{i}\mu_{i})R_{i}}) \end{eqnarray}

\begin{align} s.t.\ \sum\limits_{i=1}^{m}n_{i}=N \\
\lambda_{i}<n_{i}\mu_{i}  \\
n_{i}>\rho_{i} \end{align}

\begin{figure}[hbtp]
	\centering
	\includegraphics[scale=0.42]{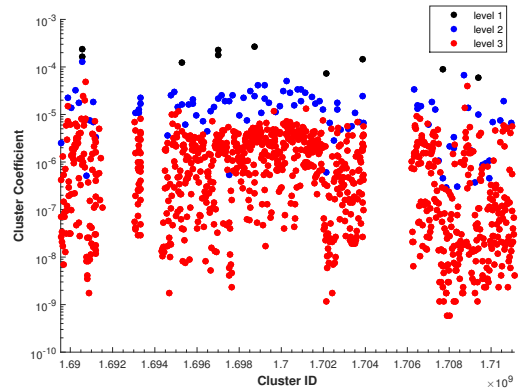}
	\caption{Cluster ID against Cluster coefficient in a log-log axis using different level of data clusters }
	\label{fig:Cluster ID against Cluster coefficient in a log-log axis using different level of data clusters}
\end{figure}

\begin{figure}[hbtp]
	\centering
	\includegraphics[scale=0.43]{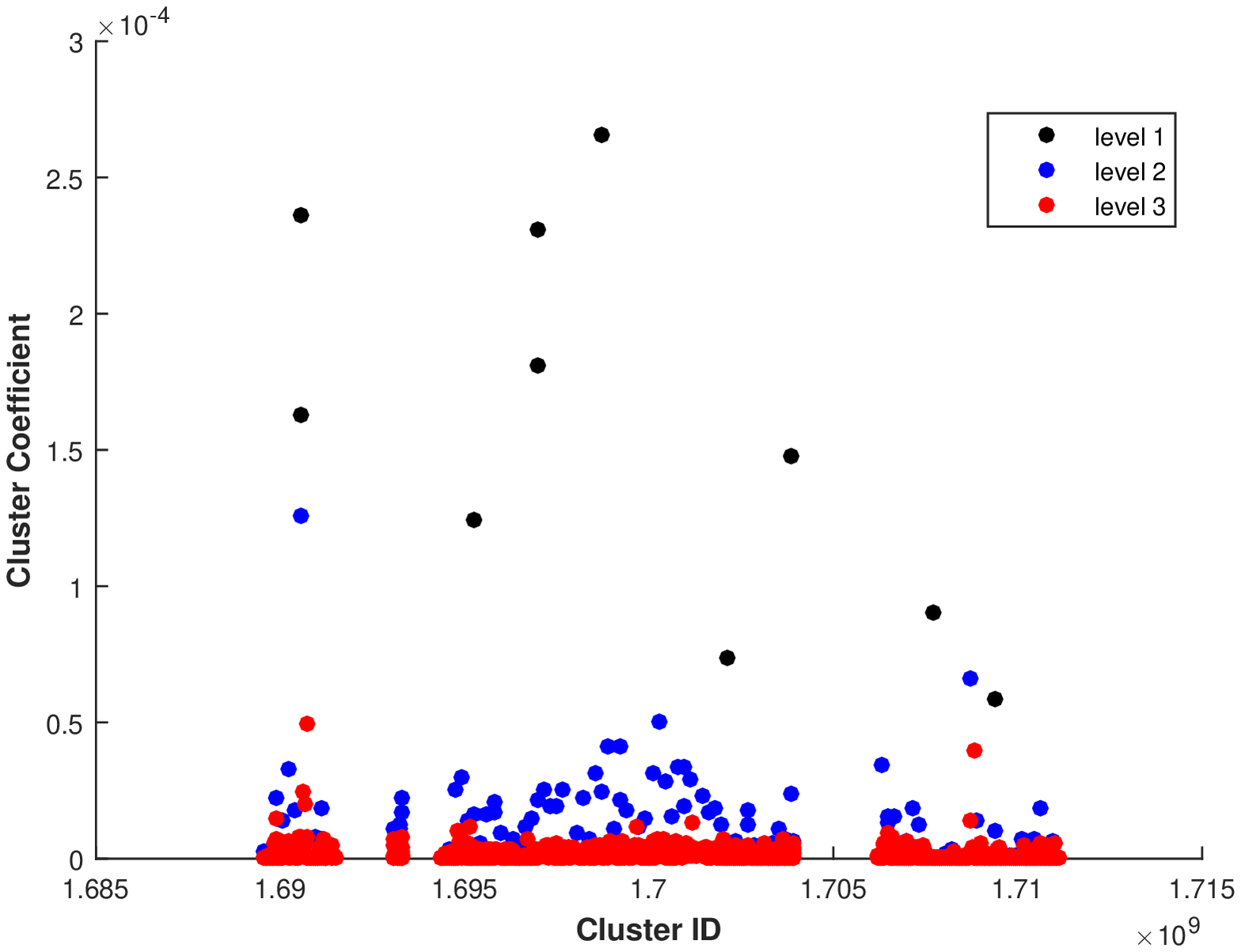}
	\caption{Cluster ID against Cluster coefficient using different level of data clusters }
	\label{fig:Cluster ID against Cluster coefficient using different level of data clusters}
\end{figure}

\begin{figure}[hbtp]
	\centering
	\includegraphics[scale=0.43]{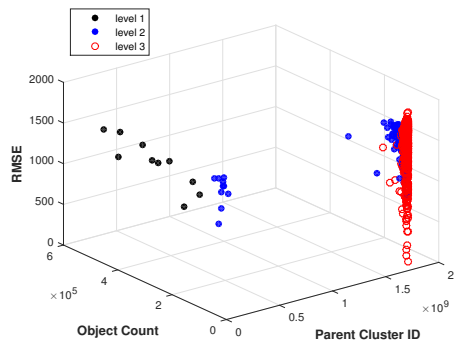}
	\caption{Model data Object, Parent Cluster ID and RMSE. }
	\label{fig:model data Object, Parent Cluster ID and RMSE.}
\end{figure}

\begin{figure}[hbtp]
	\centering
	\includegraphics[scale=0.43]{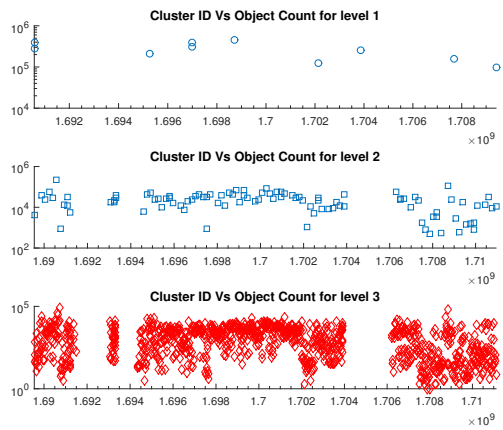}
	\caption{model data Object Count against Cluster ID for three levels of Data Statistics }
	\label{fig:model data Object Count against Cluster ID for three levels of Data Statistics}
\end{figure}

\subsection{Simulations and Experimental Results}

We have implemented our model of GPU-enabled DaaS on NVIDIA Corporation GF108GL (Quadro 600, 96 CUDA cores, 1024MB total memory) and  GF114 (GeForce GTX 560 Ti, 384 CUDA cores, 2048MB total memory ). Our data set ws model as a spatiotemporal converted data.

With Decision Support System (DSS) on Cloud systems, the collaborator or broker can facilitate a "win-win" SLA model for all data centers or cloud systems. The DaaS model will benefits connection-oriented cloud based services by acting as a collaborative plane.
This could be done by blocking their respective model of virtual routing data request especially with respect to a region \cite{7056063,5961940,7140733,Chatziantoniou:2014:IDC:2627770.2627773,Kambatla20142561,5170921}.

\begin{figure}[hbtp]
	\centering
	\includegraphics[scale=0.17]{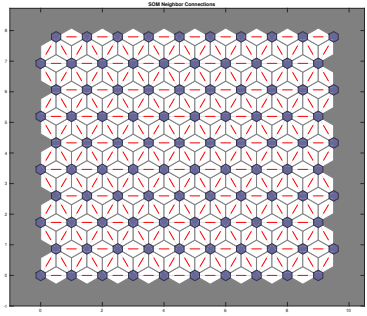}	
	\caption{SOM networks with hextop and gridtop topologies but not tritop or randtop using MATLAB Neural network computing model}
	\label{fig:SOM networks with hextop and gridtop topologies_but not tritop or randtop}
\end{figure}

The model of data that is used in our case is for agricultural based data, the database for NDVI is model along our data sets  as shown in \cite{montandon2008impact} . From Figures \ref{fig:SOM Neighbour weight distances} and \ref{fig:SOM networks with hextop and gridtop topologies_but not tritop or randtop} we have the DaaS model with SOM networks topologies to show the distribution of clustering. The learning of the different level of data is shown and it is relatively distinct. The MATLAB Neural network computing model also is used to provide a visual mode. 	The SOM model gives a good discovery of values of the clustering with respect to the DaaS. This shows that SOM can be used to cluster even DaaS models on GPU with good values. From Figure \ref{fig:Cluster ID against Cluster coefficient in a log-log axis using different level of data clusters}, Figure \ref{fig:Cluster ID against Cluster coefficient using different level of data clusters}, Figure \ref{fig:model data Object Count against Cluster ID for three levels of Data Statistics}, Figure \ref{fig:model data Object, Parent Cluster ID and RMSE.}, Figure \ref{fig:model data Object Count, RMSE against Cluster ID for high levels of Data Statistics clustering} also shows clustering of data sets with different ID and level, learning object against the RMSE value to check the error concerned and against the parent clustering model done. The results show the capability of the model to achieve a considerable better result with GPU- enabled system. Also the distribution obtained is good for our experimentation on model for DaaS.

\begin{figure}[hbtp]
	\centering
	\includegraphics[scale=0.55]{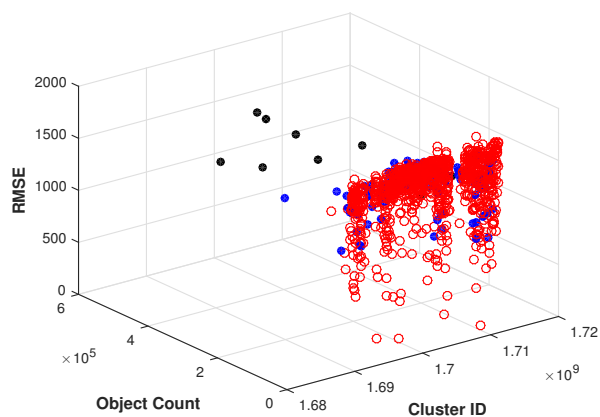}	
	\caption{model data Object Count, RMSE (root-mean-square error) against Cluster ID for high levels of Data Statistics clustering}
	\label{fig:model data Object Count, RMSE against Cluster ID for high levels of Data Statistics clustering}
\end{figure}

\section{Conclusion}

In this work, we illustrated that the use of GPU based model for data analytics in data, network centers and cloud system is with the advantage of speed especially when we offload the core operation the use the use of GPU enabled algorithm for modified k-means clustering with respect to pre-processing of data. A SOM is used to generate a unique patterns of artificial neural network for the DaaS model. Future works will deal with other different layers of data and services with respect to data that are not specific. The preliminary result shows that performance and speed can be increased with GPU based models considering a relative huge amount of data to be handled and with computation time within means. The introduction of a neural network model to learn the cluster on GPU too shows good correlation of the pre-process in order to achieve a real-time DaaS. Further work will use other supervised machine learning techniques to show the DaaS GPU based model so as to make data service a pervasive and scenario-aware through deep learning. Also, there is a need to integrate an optimization of different type of data and objective functions with modern infrastructure especially in cloud computing. With this performance increase in the GPU based systems, these cases and significant results could be achieve both to the vendor, the ISPs and the customers at large.

\section{Future Works}
The future works will focus more on the complete analytical model for GPU-enabled on cloud based networks since an exponential increase of traffic is expected to occurs for the next decade with high demand for quality-of-service (QoS) provisioning and context awareness. Further, the impact and performance analysis of energy model and respective data centers' energy consumption models will be investigated as considerably mentioned in \cite{7416982}. We will check the ability of filtering of data service based on SOM to increase the efficiency of the model.

\label{Acknowledgement}
\section{Acknowledgement}
This work is partially funded by the Turkish Agricultural Monitoring and Information System(TARBIL) and ABLNY Technologies.


\nocite{*}

\bibliographystyle{spmpsci}      
\bibliography{references}   


\end{document}